\documentclass[pre,twocolumn,showpacs,floatfix]{revtex4}
\usepackage{amsmath,amssymb,mathptm,mathrsfs}
\usepackage{color}
\usepackage{graphicx}
\usepackage{amssymb,amsfonts,amsmath}

\newcommand \bea {\begin{eqnarray}}
\newcommand \eea {\end{eqnarray}}

\makeatother

\begin{document}

\title{Intriguing effects of underlying star topology in Schelling's model with blocks}
\author{Guifeng Su, Qi Xiong and Yi Zhang}
\email{yizhang@shnu.edu.cn}
\affiliation{$^{1}$ Department of Physics, Shanghai Normal University, Shanghai 200234, P. R. China}

\begin{abstract}
We explore the intriguing effects of underlying star topological structure in the framework of
Schelling's segregation model with blocks.
The significant consequences exerted by the star topology are both theoretically analysed and numerically
simulated with and without introducing a fraction of altruistic agents, respectively.
The collective utility of the model with pure egoists alone can be optimized and the optimum stationary
state is achieved with the underlying star topology of blocks. More surprisingly, once a proportion of
altruists are introduced, the average utility gradually decreases as altruists' fraction increases. This
presents a sharp contrast to the results in Schelling's model with lattice topology of blocks.
Furthermore, an adding-link mechanism is introduced to bridge the gap
between the lattice and the star topologies, and extend our analysis to more general scenarios. A novel
scaling law of the average utility function are found for star topology of blocks.

\end{abstract}
\maketitle


\section{Introduction}

The social structures (even behaviors) share some common features with a variety of physical systems. The
collective order might emerge from simple and local social rules based on their expectations and/or decisions
among individuals.
Given the success and generality of conceptual framework of statistical physics~\cite{Path2011},
the interest of applying the methodology of statistical physics to interdisciplinary fields such as social
and economics phenomena (see e.g., \cite{Lato2007,Cast2008,Bart2019} and references therein) has grown
rapidly in recent years.

Schelling's social segregation model is probably the most celebrated example in social-economics
sciences~\cite{Sche1971,Sche1978}.
It has kept attracting attention and inspiring variant models, ranging from social
sciences \cite{Laur2003,Clar2008,Foss2009},
economics \cite{Fagi2007,Panc2007,Grau2012}, to mathematics \cite{Poll2001,Gerh2008,Rich2014}, and
statistical physics \cite{Vink2006,Stau2007,Dall2008,Gauv2009,Jensen2009,Gauv2010,Domic2011,Jensen2018}.
The segregation phase transition discovered in the model was proved to be robust in the sense that similar
outcomes occurred in variant models, even when the utility function is non-monotonic with the fraction
of similar neighbors, or the different underlying topological structures, such as lattice, random
graphs, or fractal were used~\cite{Fagi2007}. To this point, we must
emphasize that all these topological structures are simply network of grids, i.e., one agent per
site. Under such circumstances, an important observation is that, the individual agent in the model moves
to update her own utility, irrespective of the neighbors' state. It can not realize collective
optimization without nontrivial changes to the model. One of nontrivial changes, as J. C. Schelling already
stated amazingly in his classic~\cite{Sche1971}, is to introduce so-called ``blocks'' into the model, each
block contains many agents.
Unfortunately, this type of Schelling's model had been overlooked for a long time.
The exception, to the author's knowledge, is ref.~\cite{Jensen2009,Jensen2018}.
There a Schelling's model with a lattice of blocks were built, and a giant catalytic effect was found when
a small fraction of altruistic agents were introduced.
One naturally expects that more complex topologies, other than the lattice one, may have significant effects.


In current paper, we demonstrate how a specifically designed star spatial topology of blocks in Schelling's
model can achieve some novel and counter intuitive effects such like the collective optimization.
The use of the star topology is due to its ``limiting'' feature, i.e.,
the extreme heterogeneity with which the topology consists of one node with the highest possible degree (hub)
and others with only one connection (periphery), comparing to other network topologies, e.g., the
Erd\H{o}s-R\'{e}nyi (ER) random graph~\cite{Erdos1959}. One of our goals is to explore the topological
effects under such a limiting scenario.
Indeed, as one will see in the following, our theoretical analysis shows that the star spatial topology
\emph{alone} optimizes the collective utility function and drives the system into the social optimum state,
\emph{without} introducing any fraction of altruists to coordinate with dominated egoists (see, e.g.,
\cite{Jensen2018}).
However, more surprisingly, once a proportion of altruists are introduced, the optimization of the collective
utility decreases as the altruists' fraction increases. The numerical simulations fully confirm our predictions.
This forms strong contrast to the results of lattice topology.

Not only did we analyze and simulate the scenario of the limiting star topology,
but also that of more general ones. This is done by randomly adding links between any pair of peripheral blocks.
This adding-link mechanism bridges the gap between the lattice and star configurations, since the former (lattice)
corresponds to a fully-connected network. We can probe the boundaries of the effects in terms of the mechanism,
and hence extend our analysis to more general topologies.


The rest of the paper is organized as follows: in Sec.~\ref{sec:II}, we define the Schelling's model on a
star spatial topology of blocks. In Sec.~\ref{sec:III} we present our major analytical results and numerical
simulations.
Firstly we show the evolutionary dynamics of the Schelling's model with star topology of blocks in detail,
and the existence of the optimum stationary state is demonstrated; then for the stationary state we show
theoretically that with a fraction of altruists being introduced, the collective
utility function rather decreases surprisingly.
Further in Sec.~\ref{sec:IV}, in order to explore the boundary of the effects due to the star topology of blocks,
we consider the adding-link mechanism, which bridges the gap between two topologies.
The mechanism leads to some striking results on the optimum state of the system.
Finally, we conclude the paper with the summary of our findings in Sec. ~\ref{sec:V}.

\section{The Schelling's Model with Star Topology of Blocks}
\label{sec:II}

\begin{figure}[ht!]%
\centering
\includegraphics[width=0.9\columnwidth]{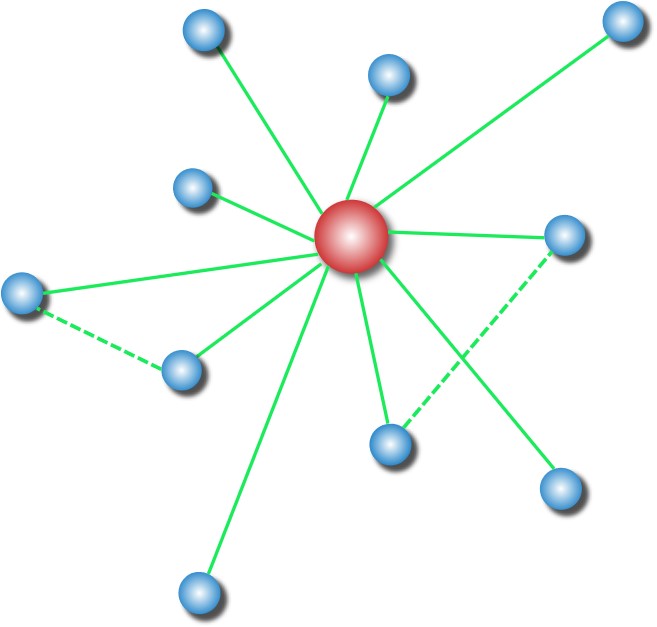}%
\caption{(Color online) A cartoon illustration of a ``city'' with star spatial topology. It can be viewed
as an abstraction of a monocentric city with satellite blocks. A total number of $N$ agents live in the city.
The blue (peripheral grey) nodes represent the peripheral blocks, the red (central dark) node represents
the central block, so called ``hub'' (an analogy of central business district in real world),
who owns the most number of edges. The solid lines represent the links of all neighboring blocks to the central
hub. The dashed lines represent potentially possible
links among two peripheral blocks (see Sec.~\ref{sec:IV}).}
\label{fig1}%
\end{figure}

Consider a total number of $N$ agents living in a city with star topology of blocks (see Fig.~\ref{fig1}
for a cartoon illustration of this star configuration).
From social-economical perspective, the star topology can be visualized as a mono-centric city with some
satellite blocks (e.g., a central business district can be the hub and all agents from different blocks
commute there).
Note that the peripheral blocks are all connected with the central hub but no links between one another.
The city is homogeneously divided into $Q \gg 1$ non-overlapping blocks, and each block into $H$ sites with
the capacity of a maximum $H$ agents.
Initially, a $N = Q H \rho_0$ agents are randomly distributed over the blocks, the initial average block
density is $\rho_0$, and in general $\rho_0 < 1/2$.
One has to know the specific form of the utility function in order to obtain the equilibrium
configurations. We assign all agents the same utility function $u(\rho)$, which depends on the agents
density $\rho$ of the block they live in. It takes a symmetric triangular form:
\begin{equation}
u(\rho) = \left\{
\begin{array}{rl}
2\rho,\, & \text{if } \rho \leq 0.5,\\
2(1-\rho),\, & \text{otherwise} .
\end{array} \right.
\end{equation}
One defines the collective utility $U$ as the sum of all agents' utilities,
\begin{equation}
U = H \sum_{q=1}^Q \rho_q u(\rho_q) \,,
\end{equation}
and the average utility per agent is simply $\langle u\rangle = U/N$. The density of agents in
the $q$-th block is $\rho_q = n_q/H$ ($q = 1$, $\cdots$, $Q$). We already assumed that each site contain at
most one agent, so that the number of agents in the $q$-th block, $n_q$, satisfies $n_q \leq H$.
The utility $u(\rho_q)$ that each agent shares describes the degree of satisfaction concerning
the density of the block the agent is located.

According to the spirit of Schelling's model one can introduce two types of agents: ``egoists'' and
``altruists'' depending on one's need. The former (egoist) acts to improve her/his individual utility,
and the latter (altruist) to improve the collective utility. Egoists consider variation of their
individual utility $\Delta u$, while altruists consider the variation of the global utility $\Delta U$.
At each time step, one randomly selects an agent and a vacant site in another connected block. The agent
accepts to move to this new site only if its objective function strictly increases, and the moving agent is
taken into account to compute the density of the new block. Otherwise, it stays in its present block. Then,
another agent and another empty site are chosen at random, and the same process repeat until a stationary
state is reached, in which there are no possible moves for any agent. In general, one can apply or extend
the framework of equilibrium statistical mechanics to the Schelling's model in order to find the stationary
probability distribution for the microscopic configurations~\cite{Callen1985}. However, from an individualistic
point of view, one has to resort to numerical simulations in order to obtain general solutions.

\section{Results and Discussions}
\label{sec:III}

\subsection{Evolutionary dynamics}

One of the most intriguing results of the star topology in Schelling's model is that an optimum stationary
state can be obtained with egoists alone. To see how the system approaches its optimum stationary state,
it is necessary to address the evolutionary dynamics of Schelling's model with given star topology.

Initially, there are $N \equiv \rho_0 QH$ egoists randomly distributed over all blocks. An initial total
average density of the agents $\rho_0=0.4$ is taken throughout the paper.
The relaxation process to the stationary state can be divided into the following three stages:

(1) at first,
egoists from peripheral blocks immigrate to the hub block gradually, as long as the utility of the hub
$u_{\rm hub}(\rho_{\rm hub})$ is not the lowest one at the very beginning. Without losing generality, suppose
that there are $Q_l>0$ blocks of which utilities $u_l(\rho_l>0)$ are lower than that of the hub block. The
inflow rate of the agents to the hub, $P_{\rm in}=Q_l/Q$, is larger than the outflow rate
$P_{\rm out}=(Q-1-Q_l)/(Q(Q-1))$ of the agents, given that the probability for an arbitrary block in the
topology to be chosen is $1/Q$. With the increasing of the hub density $\rho_{\rm hub}$ through the aggregation
of agents, the utility $u_{\rm hub}(\rho_{\rm hub})$ is optimized
at first when $\rho_{\rm hub}<1/2$ and turn to decreased when $\rho_{\rm hub}>1/2$;

(2) Similar to
$u_{\rm hub}$, as $\rho_{\rm hub}$ increases and approaches to $1/2$ (i.e. $\rho_{\rm hub}\nearrow 1/2$),
$Q_l \rightarrow Q-1$, and then decreases gradually until $Q_l=1$.
This procedure dominates the global dynamical process, and can be approximately considered as
Schelling's model with only two blocks and pure egoists. In this case the statistical physics method can be
directly applied~\cite{Jensen2018}. Accordingly, the egoists tend to maximize a global parameter
``effective free energy'' $L$ which is defined as $L=\sum_q l_q$, where $l_q$ is the free energy of the $q$-th
block, and can be approximately written as $l_q(\rho_q)\approx H\int_0^{\rho_q}u(\rho)d\rho$, for block
$q=1,2,\cdots,Q$ in the large $H \gg 1$ limit.
Depending on the initial density $\rho_0$, the optimization requirement of $L$ in a two blocks system leads to
two kinds of different phases, one for the average density $\bar{\rho}\leq 1/2$ where the egoists totally
gathered at one block and leaving another empty, the other for $\bar{\rho} >1/2$ where the egoists equally
distributed over both blocks. Recall that the initial total average density $\rho_0 < 1/2$ is taken, and only
the remaining blocks with $u_l(\rho_l < 1/2) < u_{\rm hub}(\rho>1/2)$ will be exhausted and the hub eventually
extracts all egoists from them. After this node-pair effect stage that leads to $Q_l=0$, hence $P_{\rm in}=0$,
the egoists gathered at hub block start to equally hop to non-empty peripheral blocks. The utilities $u(\rho_q>0)$
of these blocks, including $u_{\rm hub}(\rho_{\rm hub})$, increase again, and are accompanied with the quickly
decreasing of $\rho_{\rm hub}>1/2$ and slowly increasing of $\rho$ in other blocks with $u(0<\rho <1/2)$.
Once other peripheral blocks with $u_l(\rho_l < 1/2)< u_{\rm hub}(\rho>1/2)$ reappeared, i.e. $Q_l=1$, the
node-pair effect occurs again and takes the leading role in global dynamics. Through such reoccurrences during
the evolution, the hub persistently relocates the egoists from an arbitrary peripheral block with
$u_l(\rho_l < 1/2)< u_{\rm hub}(\rho>1/2)$ to other peripheral blocks with $\rho>0$, whose utilities will be
improved continuously;

(3) The prevalence of the same node-pair effect will last until $\rho\ge 1/2$ is achieved
for each occupied peripheral blocks.
This suggests $\bar{\rho} \gtrsim 1/2$ for all remaining $\rho>0$ peripheral blocks.
Therefore, existing blocks will no longer be exhausted, since the phase for $\bar{\rho} \gtrsim 1/2$
takes over the node-pair effect, and results in an equally distribution of egoists over the remaining
blocks. Finally, a stationary state is approached with an optimized average utility
$\langle u(\rho \gtrsim 1/2) \rangle \lesssim 1$.

On the other hand, in contrast to the pure egoists' case, the behavior of altruists can lead to a totally
different, more surprising phase, where a stationary state with lower $\langle u \rangle$ will be achieved.
Since their target function is to improve the collective utility $U=\sum u$, the altruists from peripheral
sites cannot optimize $U$ by freely hopping to other sites, but only judge whether moving to the hub.
However, the density of the hub $\rho_{\rm hub}>1/2$ always remains the largest one during the evolution,
as long as it is not too small in initial stage. Thus for altruists the only way moving out peripheral
sites is prohibited by the high density of the hub, otherwise the total utility $U$ would be lowered down.
Hence for the case of pure altruists (i.e. $p=1$), almost all agents will stay at their initial sites without
any moves, except the initial aggregation to the hub block until $\rho_{\rm hub}=1/2$ is reached.

For general coexistence case, e.g., $p=0.5$, the egoists keep moving and optimizing their own utilities $u$
and ultimately optimize of the collective utility $U$. At the same time, prohibited by the high density of
$\rho_{\rm hub}>1/2$, most altruists from peripheral blocks have to stay at their initial sites and can not
improve the collective utility $U$. Recall that the density of the hub $\rho_{\rm hub}>1/2$ always remain
the largest one during most time of the evolution, as a result, the rest altruists located at the hub
initially would escape from the hub and randomly migrate to any other peripheral blocks.

\begin{figure}[ht!]%
\centering
\includegraphics[width=\columnwidth]{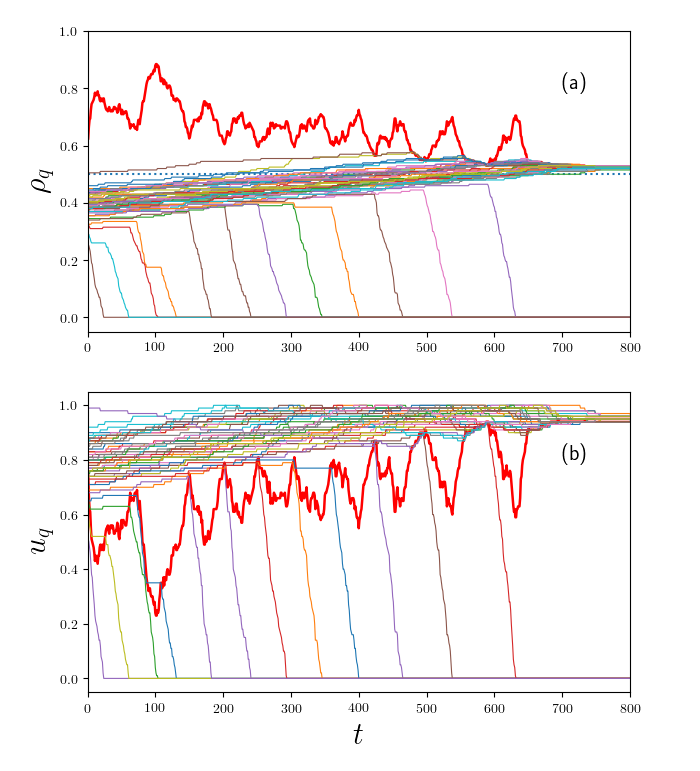}
\caption{(Color online) (a) Evolution of agents' density $\rho_q$, and (b) the utility $u_q$, of the $q$-th
    block of star spatial topology with $H=200$ and $Q=50$. Note that an altruists' fraction $p=0$ is taken in
    the simulation, which means no altruists is introduced. The red solid (bold dark) line represents the
    evolution of hub's density $\rho_{\rm hub}$ in (a) and the utility $u_{\rm hub}$ in (b), respectively.
    A horizontal blue (grey) dotted line in (a) at $\rho_q=0.5$ is for eye guidance.}
\label{fig2}
\end{figure}

\begin{figure}[ht!]%
\centering
\includegraphics[width=\columnwidth]{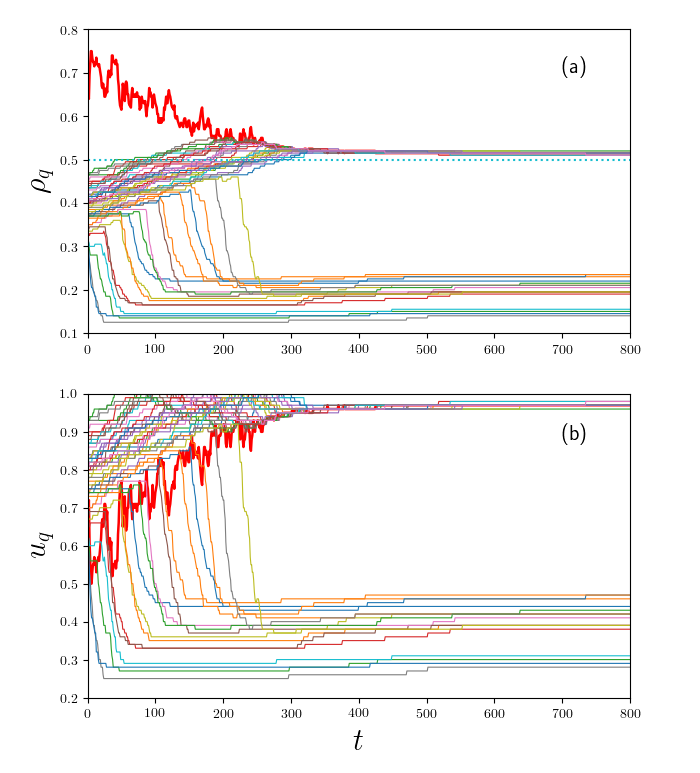}
\caption{(Color online) Similar to Fig.~\ref{fig2}, but with an altruists' fraction $p=0.5$.
   }
\label{fig3}
\end{figure}

\begin{figure}[ht!]%
\centering
\includegraphics[width=\columnwidth]{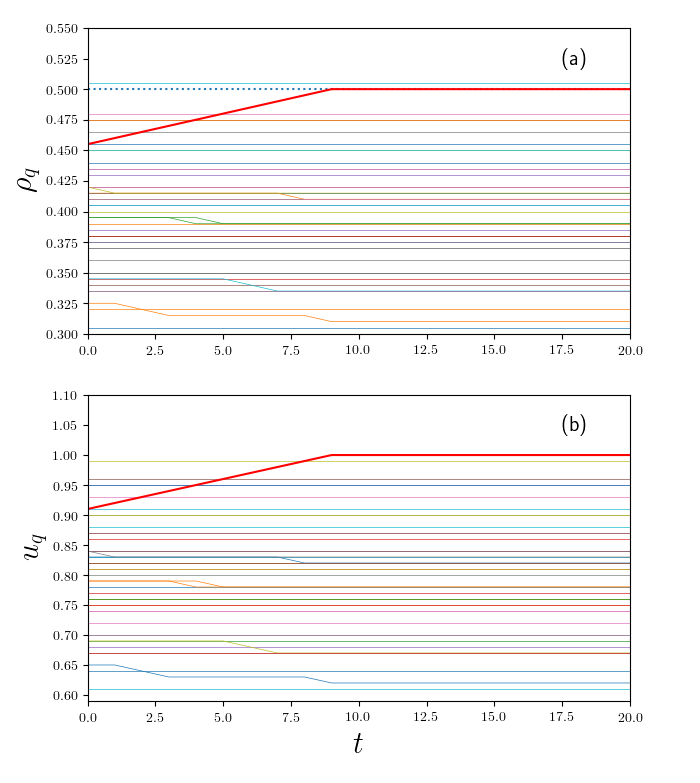}
    \caption{(Color online) Similar to Fig.~\ref{fig2}, but with an altruists' fraction $p=1$.
    }
	\label{fig4}
\end{figure}

We demonstrate the above scenario of dynamical evolution in Fig.~\ref{fig2}, Fig.~\ref{fig3}, and
Fig.~\ref{fig4} with fixed $H=200$ and $Q=50$, but different altruist's fractions $p=0$, $p=0.5$, and $p=1$
respectively. In Fig.~\ref{fig2}, we show the evolution of agents' density $\rho_q$ (upper panel), and
utility $u_q$ (lower panel), of the $q$-th block of star spatial topology with an altruists' fraction $p=0$.
Each one of the curves represent a $\rho_q$(upper panel) and a $u_q$ (lower panel) of a $q$-th block. Note
that with the dropping of each $\rho_l$ (or $u_l$), displayed by those curves which sharp declining to zero,
$\rho_{\rm hub}$ (the bold dark line in the upper panel) increase while $u_{\rm hub}$ (the bold dark line
in the lower panel) decrease until $\rho_l\rightarrow 0$ (or $u_l\rightarrow 0$).
On the contrary, $\rho_{\rm hub}$ decreases while $u_{\rm hub}$ increases until the appearance of another
$\rho_l$ (or $u_l$). The process repeats until all $\rho_q>0$ converge gradually to a value slightly larger
than $0.5$, and all $u_q>0$ almost approach unity finally.

The similar process occurs in Fig.~\ref{fig3}, except for $p=0.5$. One significant difference comparing
to Fig.~\ref{fig2} can be observed: almost all $\rho_l$ (or $u_l$) would not get down to zero since the
altruists still stay at those blocks. Dragging down by these small but nonzero $u_l$'s, the average utility
$\langle u\rangle$ is inversely proportional to the altruists' fraction $p$.

Likewise, the process with $p=1$ is plotted in Fig.~\ref{fig4}. Almost all agents will stay at their initial
site without any moves, except the initial aggregation to the hub block until $\rho_{\rm hub}=1/2$ is reached,
which means the average utility $\langle u\rangle$ is not optimized at all (see the green dashed lines in
upper panel for $\rho_q$ in Fig.~\ref{fig4}).

\subsection{The optimum stationary state}

Based on the theoretical analysis and numerical simulations for evolutionary dynamics above, the optimum
stationary state as a function of altruists' fraction $p$ can be calculated directly
The analytical results of the optimized stationary state for $p=0$ case can be obtained from a recurrence
relation $\rho_n=\rho_{n-1}+\rho_{n-1}/(Q-n)$, where $\rho_n$ represents the average density of occupied
blocks when there are $n$ empty blocks. Given that $\rho_1=\rho_0+\rho_0/(Q-1)$, the density of occupied
blocks (i.e., $n^\ast$ empty blocks) at stationary state is approximately
\begin{eqnarray}
\label{eqnn}
\rho_{n^\ast} &\approx& \rho_0 \Big(1+\frac{1}{Q-1}\Big) \Big( 1+\frac{1}{Q-2}\Big) \ldots \Big( 1+\frac{1}{Q-n^\ast}\Big) , \nonumber \\
              &=& \rho_0 \Big(\frac{Q}{Q-n^\ast}\Big) ~.
\end{eqnarray}
Recall that when $\rho \geq 1/2$ is satisfied for all occupied blocks, the node-pair effect disappears.
For optimum stationary state, Eq.(\ref{eqnn}) can be written as
\begin{equation}
\rho_{n^\ast}=\rho_0\Big(\frac{Q}{Q-n^\ast}\Big) = \frac{1}{2}+\epsilon ~,
\end{equation}
where $\epsilon$ is a small quantity induced by fluctuation. Obviously, the
number of occupied blocks at stationary state satisfy $Q-n^\ast =\text{Int}[ 2\rho_0 Q-\epsilon ]$, where
$\text{Int}[\cdots]$ represents the maximum integer of $[ \cdots ]$, and hence the analytical results of
density $\rho_{n^\ast} = 1/2+\epsilon$. The collective utility is then $\langle u\rangle=2(1-\rho_{n^\ast})$
for the optimum stationary state.

In coexistence situation $0<p<1$ in stationary state, the hub block is occupied by egoists and low
density blocks by altruists only, the optimized peripheral blocks are occupied by both egoists and
altruists.
One expects that the average utility $\langle u(p) \rangle$ (and the total utility $U$) may decrease
as $p$ increases. In the limit of pure altruists' case $p=1$, since the hardly moving appearance of
altruists in star topology, the stationary density of each peripheral block $\rho_q$ would thus
almost keep unchanged in the $Q\rightarrow \infty$ limit.
It is easy to found that $\sum_{q=1}^{Q-1} \rho_q \approx Q\rho_0$, hence the average utility
$\langle u\rangle \simeq u(\rho_0)$ holds for all peripheral sites.

\begin{figure}[ht!]%
\centering
\includegraphics[width=\columnwidth]{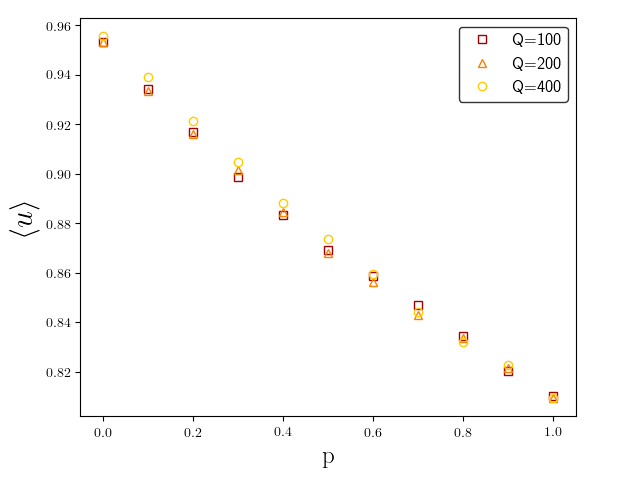}
\caption{(Color online) The average utility function $\langle u \rangle$ as a function of the altruists'
         fraction $p$ with $H=200$, and $Q=100$ (open squares), $Q=200$ (open triangles), and
         $Q=400$ (open circles), respectively.}
\label{fig5}
\end{figure}

The numerical simulation results fully agrees with our theoretical analysis.
In Fig.~\ref{fig5}, we show the average utility $\langle u(p) \rangle$ versus the parameter $p$ for
different number of blocks, $Q=100$, $Q=200$, and $Q=400$, respectively. For simplicity, throughout
our numerical simulations, we fix the parameters $H=200$ and $\rho_0=0.4$. We have three observations
from the figure, as expected from theoretical analysis:
(1) the system approaches its optimum state even without introducing any altruists (see the left most
$p=0$). Note that it is \emph{impossible} to emerge such an optimum state in Schelling's model on a
lattice city with pure egoists~\cite{Jensen2018};
(2) indeed, the average utility $\langle u(p)\rangle$ (and/or the total utility $U$) \emph{decreases}
linearly as the altruists' fraction $p$ increases, and this is also a surprising consequence since
in \cite{Jensen2018}, an infinitesimal small fraction $p \sim 1/Q$ can drive the system into the optimum
state abruptly. Increasing the fraction $p$ will not change the situation;
and (3) the average utility $\langle u\rangle$ is independent of the number of blocks $Q$, and computed
collective utility data for $Q=100$, $200$, and $400$ collapse onto the the same curve nicely.

\section{Adding-link Mechanism in Schelling's Model with Star Topology of Blocks}
\label{sec:IV}

So far we have demonstrated the significant roles played by underlying star spatial topology of blocks in
Schelling's model, both on its optimum stationary state and evolutionary dynamics. On the other hand, the
mechanisms of racial/social segregation with relocation of agents in city naturally depends on spatial
proximity between blocks of agents. The extreme heterogeneity of the star topology represents a limiting
scenario of relevant effects. One may be wondering to what extent such topological effects present. We try
to answer this question in this section by introducing so-called adding-link mechanism.

From perspective of complex network, the regular lattice corresponds to a
fully-connected network among blocks. Based on the star topology, one can randomly adding links to
each pair of \emph{peripheral} blocks of star configuration with some adding-link probability $\Pi \in [0, 1]$,
as illustrated in Fig.~\ref{fig1} (see green dashed lines for illustration).
This mechanism bridges the gap of two topologies in the present model, so that one underlying topology can
evolve towards the other, or vice versa, as the adding-link probability varies.
This sets two limits: one is $\Pi \rightarrow 0$, which corresponds to the star topology;
the other one is $\Pi \rightarrow 1$, which gives the fully-connected network topology.
By adjusting this adding-link probability, one can observe how the system is driven towards the optimal state,
i.e., the collective utility $\langle u \rangle = 1$, and hence has the chance to study the transition of the
corresponding effects under different topologies, and the limit of results in \cite{Jensen2018}.

The theoretical aspect of this transition is rather straightforward. Although the star topology is responsible
for the optimization of pure egoists, adding more connections between any pair of peripheral blocks may destroy
the global optimal state.
One expects that the global optimization of egoists would be failed when the whole peripheral blocks are
connected by one new link on average. This is due to the following physical picture.
Considering $Q-1$ peripheral blocks (excluding the central hub) in the star topology, and they are randomly
connected with a probability $\Pi$. This is exactly the procedure to form an ER random graph~\cite{Erdos1959}.
For an ER random graph with $Q-1$ nodes, there exists a critical probability $\Pi_c \propto 1/(Q-1) \sim 1/Q$
($Q \gg 1$), or a critical average degree $\langle k\rangle_c=1$ for the emergence of the largest cluster.
The critical value of $\Pi_c$ corresponds to an abrupt change in underlying structure, and hence leads to
a complete destruction of the optimization. A simple scaling analysis gives us a scaling of the averaged
utility function versus re-scaled connecting probability $\Pi^{\ast} = Q (\Pi-\Pi_c)$.

\begin{figure}%
\centering
\includegraphics[width=\columnwidth]{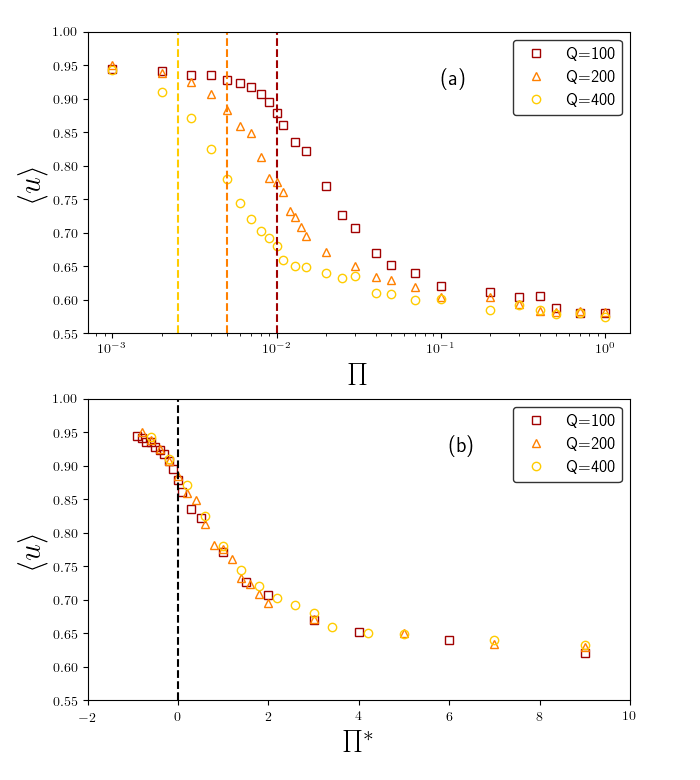}
 \caption{(Color online) (a) The average utility $\langle u \rangle$ as a function of adding-link probability $\Pi$
       (note the linear-log scale) for different numbers of blocks, $Q=100$ (upper open squares), $Q=200$
       (middle open triangles), $Q=400$ (lower open circles), respectively, with a fraction of altruists $p=0$.
       Three vertical dashed lines correspond to the critical probability $\Pi_c$'s for $Q=100$ (right),
       $Q=200$ (middle), and $Q=400$ (left).
       (b) Collapse of $\langle u \rangle$ after re-scaling with respect to $\Pi^{\ast} = Q(\Pi-\Pi_c)$
       (in linear scale), using the same data as in (a).
       The dashed line labels the re-scaled critical probability $\Pi^{\ast}_c$.}
\label{fig6}
\end{figure}

The above theoretical analysis are verified by numerical simulations and the major results are summarized
in Fig.~\ref{fig6}. From the upper panel of Fig.~\ref{fig6}, for different numbers of blocks, $Q=100$
(open squares), $Q=200$ (open triangles), and $Q=400$ (open circles), respectively, one sees that
as the probability $\Pi$ increases, the optimization of the utility function induced by this star topology
fades away. When $\Pi$ reaches unity ($\Pi \rightarrow 1$), a fully-connected underlying topology is formed,
and it is topologically
equivalent to the lattice city used in most literature, e.g., \cite{Sche1971,Jensen2018}, to name a few.
However, the expectation value of utility function $\langle u(\Pi) \rangle$ is far less than the optimized
one.
In $\Pi \rightarrow 0$ limit, one has the star topology and the optimized utility is approximately achieved.
While in between, i.e., $0 < \Pi < 1$, the effects present themselves for more general underlying topologies.
They are mainly due to two factors of the underlying topological structure according to previous analysis:
the mono-centric hub in the topology, and the connections between peripheral blocks. This suggests that these
effects hold for more general topologies consisted of a core network of arbitrary topology, e.g., a
fully-connected one or a random graph, plus a finite proportion of peripheral blocks, provided the portion of
the core network is small enough. Even for an over-simplified core of one
dimensional loop, the numerical
simulations show that the same effect presents, although they are not realistic topologies anymore for real
world cities.

The other interesting discovery, i.e., the scaling of the average utility function versus re-scaled adding-link
probability $\Pi^{\ast}=Q(\Pi-\Pi_c)$ is shown in the lower panel of Fig.~\ref{fig6}. One sees clearly that the
average utility function $\langle u\rangle$ for different numbers of blocks, $Q=100$, $Q=200$, and $Q=400$,
respectively, now collapses on the same curve.

\section{Summary}
\label{sec:V}

Although it has been almost fifty years since the famous Schelling's model was proposed, it still
inspires new ideas and/or variant models in, but not limited to, social-economics and/or statistical
physics.
However, it is worth to notice that most variants of Schelling's model are based on
the lattice and/or grid network topologies, much less are based on the topology of blocks, and they
have long been overlooked. On the other hand, one can not realize the collective
optimization with underlying lattice topology, no matter it is grid-based or block-based. While for
Schelling's model with more complex underlying topologies of blocks, their social-economics and
statistical physics are still waiting for further exploration.

In this paper, we investigate the Schelling's model with underlying star topology of blocks.
The theoretical analysis of the corresponding roles played by the star topology and its hub in evolutionary
dynamics and optimum stationary state is developed.
Comparing to the common-used lattice topology, for the first time, we achieve a collective optimization
via implementing an underlying star topology. We also discover that even with pure egoists ($p=0$), the
collective utility of the model can be optimized, and the system tends to an optimum stationary state.
Physically, this means for pure egoists, an underlying spatial topology formed by these agents may help
them reach an optimum stationary state.
We emphasize that this is a purely spatial topological effect in Schelling's model with blocks.

A more surprising result is that once a proportion of altruists, i.e., $p \neq 0$, are introduced in
Schelling's model with star topology of blocks, the optimization of the collective utility
\emph{decreases} as the altruists' fraction $p$ increases. In other words, introducing more and more
altruistic agents does not help the system approaching its optimal state. This forms a sharp contrast
to previous study~\cite{Jensen2018}, in which an infinitesimal fraction of altruists $p \sim 1/Q$
produces a dramatic catalytic effect and drives the system to the optimal steady state.
The numerical simulations for evolutionary dynamics in Schelling's model with $p=0$, $p=0.5$, and
$p=1$ respectively, are fulfilled, and fully support our theoretical analysis.

The extreme heterogeneity of the star topology represents a limiting scenario. To explore the limit(s) of
the relevant effects, or in other words, to what extent such topological effects present comparing to the
lattice topology, we introduce the adding-link mechanism. As a result, the system can evolve from one
topology to the other by adjusting the adding-link probability. We conclude that as the probability
increases, the optimization of the utility function induced by the star topology fades away. When it
reaches unity, the expectation value of utility function $\langle u \rangle$ is far less than the
optimized one. When the probability goes to zero, the optimized utility is approximately achieved via
star topology.
The mono-centric hub of the topology, and the connections between peripheral blocks play the essential
role in this procedure. A nice scaling of the average utility function versus re-scaled probability is
found as well.

The social-economic meanings of the current study are still waiting for
a sufficient understanding. For instance, whether or not this provide some new insight on human
cooperations remains to be answered. We leave this to our future investigations.

\textbf{Acknowledgments} The authors Y.Z. and G.S. thank Prof. Pablo Jensen for helpful discussions. We also
thank an anonymous referee for helpful comments and giving us the opportunity to improve our manuscript.
This work is supported by the National Natural Science Foundation of China (NSFC) via Grant No. 11505115.


{}

\end{document}